# COVID-19 spreading patterns in family clusters reveal gender roles in China

Jingyi Liao[1], Xiao Fan Liu[2], Xiao-Ke Xu[3], Tao Zhou[1]
1. Big Data Research Center, University of Electronic Science and Technology of China, Chengdu, China
2. Web Mining Laboratory, Department of Media and Communication, City University of Hong Kong, Hong Kong SAR, China
3. College of Information and Communication Engineering, Dalian Minzu University, Dalian, China

**Abstract:** Unfolding different gender roles is preceding the efforts to reduce gender inequality. This paper analyzes COVID-19 family clusters outside Hubei Province in mainland China during the 2020 outbreak, revealing significant differences in spreading patterns across gender and family roles. Results show that men are more likely to be the imported cases of a family cluster, and women are more likely to be infected within the family. This finding provides new supportive evidence of the "men as breadwinner and women as homemaker" (MBWH) gender roles in China. Further analyses reveal that the MBWH pattern is stronger in eastern than in western China, stronger for younger than for elder people. This paper offers not only valuable references for formulating gender-differentiated epidemic prevention policies but also an exemplification for studying group differences in similar scenarios.

## 1. Introduction

Epidemics were largely affected by social factors, and have brought long-standing consequences to our society [1][2]. Recent studies revealed both the gender impact on COVID-19 outcomes [3] and the impact of COVID-19 on gender equality [4]. Wenham *et al.* [5] reported that although the numbers of male and female cases are equal, more men than women become critically ill patients. Adam [6] showed that women's participation in the workforce is correlated to women's share of COVID-19 deaths, which to some extent explained the women's smaller share of death from COVID-19. Vegt and Kleinberg [7] showed that during the COVID-19 pandemic, women worried more about their loved ones and severe health concerns while men were more occupied with effects on the economy and society. Liu *et al.* [8] found that male users were more inclined towards using regular textual language with fewer emojis after the pandemic, suggesting that during public crises, male groups appeared to control their emotional display. Cui *et al.* [9] empirically demonstrated that the impacts of COVID-19 on research productivity are gender different, for example. In the 10 weeks after the lockdown in the United States, although total research productivity increased by 35 percent, female academics' productivity dropped by 13.2 percent relative to that of male academics. Oreffice and Quintana-Domeque [10] investigated gender differences after 3 months of the first UK lockdown of March 2020, and found that women's mental health was worse than men: women were more concerned about getting and spreading the virus, and women perceived the virus as more prevalent and lethal than men did. In despite of the above-mentioned influential works, how gender-differentiated behaviors affect the spreading of COVID-19 and what insights can be obtained from gender-differentiated spreading patterns have not yet been fully discussed.

One typical gender role across cultures assumes men as the breadwinners and women as the homemakers (MBWH). According to the World Bank, the average employment rates of men and women in 2021 are 67% and 44% [11]. At the same time, women spend two to ten times more time on unpaid care work than men [12], specifically, women devote 2.3, 8.6, 4.8, and 1.5 times more to housework than men in China, India, Japan, and America, respectively [13]. Consequently, the average commuting distance and commuting time of men are longer than women in both developed areas like Europe [14] and developing areas like Asia [15], Latin America [16], and Africa [17]. The longer time spent in public and workplaces is associated with a higher infection risk by infectious diseases and hence a higher risk of bringing the disease back home. Some preliminary studies indicated that women have a higher risk of being infected by family members [16] or becoming infectees in community clusters [18].

We quantify the gender roles' effects on epidemic spreading in a unique setting. In December 2019, COVID-19 outbroke in Wuhan, Hubei, China. The outbreak coincided with the world's largest short-term population migration, *Chunyun,* that is, during Chinese New Year, most migrant workers and family members who live separately would return to their homes and reunite with their families. *Chunyun* accelerated COVID-19 spreading for a short period [19,20] before the urgent and strict control measures starting in late January, such as city lockdowns and house quarantine. Those control measures quickly stopped long-haul disease transmission and confined it in local communities or households [21,22]. *Chunyun* and lockdowns shaped atomic disease transmission clusters, in which the migrant workers flowing out of Hubei province started the subsequent transmissions within families. By applying gender filters to the imported and subsequent infections, we can clearly portray the gender-differentiated spreading patterns of COVID-19.

We use curated epidemiological survey data with fine-grained features [23] and reconstruct the transmission chains during the Chinese New Year (January 20 to February 18, 2020). Then, we apply statistical tests and construct null models to test the following two hypotheses.

**Hypothesis 1**: *gender role in China conforms to "men as breadwinner", and thus men have a higher probability of becoming the imported cases of the family clusters.*

**Hypothesis 2**: *gender role in China conforms to "women as homemaker", and thus women have a higher probability of being infectees in the family clusters, even after the control of the gender difference in susceptible population brought by the imbalance in imported cases.*

## 2. Data and methods

We collected the epidemiological survey reports of COVID-19 cases from 264 municipal governments outside Hubei province in mainland China. Data include cases' demographic characteristics, social relations of the close contacts, travel and contact diaries, and symptoms timelines [23]. This data set covers 12,667 confirmed cases from January 19 to November 20, 2020, accounting for 71.1% of all confirmed cases outside Hubei Province. This paper focuses on the cases during the Spring Festival (January 20 to February 18, 2020), including a total of 7,261 confirmed cases (3,822 men, 3,367 women and 72 unknown). The number of confirmed men is significantly higher than confirmed women in the early stage of the outbreak. With the development of the epidemic, the difference gradually reduces (see Fig. 1).

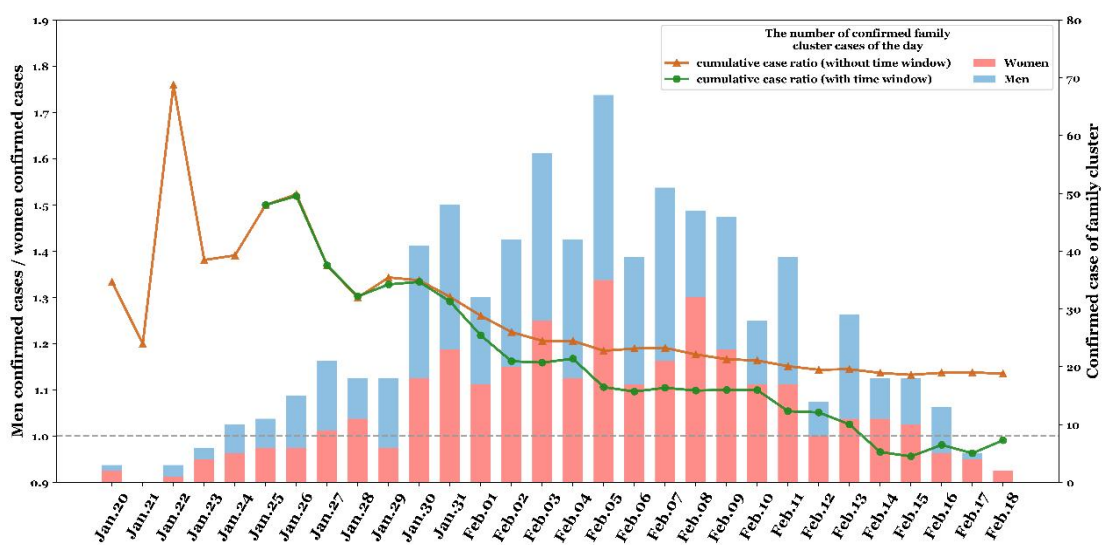

**Figure 1.** The number of daily confirmed cases in family clusters and the ratio of confirmed men to women. The red and blue bars represent the number of confirmed women and men, respectively. The cumulative men-to-women ratios are represented by the two curves, with the black one representing the cumulative ratio in a 7-day time window up to the reported date and the grey one representing the cumulative ratio up to the reported date.

We follow the procedure of [24] to reconstruct transmission chains from the epidemiological survey reports. Potential transmission pairs are first identified from close contacted confirmed cases. The infector is the one with a travel history and explicit locations of exposure. When a case has multiple potential infectors, the one with the earliest symptom onset time is chosen as the infector. Of the 7,261 confirmed cases, 1,108 have at least one infected family member (here, a family member is the one with lineal consanguinity within three generations). We label these cases as family-clustered cases. This paper does not treat siblings as family members because grown siblings live separately in China, and most children under 18 have no siblings for the past 30 years, attributed to the one-child policy. For convenience, if A infects B in a family cluster case, we call A the infector and B the infectee. Notice that, a family member may be both an infector and an infectee. For example, if A infects B and B infects C, then B is both an infector and an infectee. As a result, in the later statistics, the number of infectors is larger than the number of imported cases.

## 3. Results

### 3.1 Gender-differentiated transmission risks

The 1,108 cases form 446 family clusters (see Fig. 2). Among them, 300 have a size of 2, 99 are of size 3, 47 are of size larger than 3, and the largest cluster size is 8. Table 1 shows the numbers and proportions of all clusters with sizes 2 and 3. It is evident that the proportion of the men → women transmissions is significantly higher than the inverse women → men transmissions.

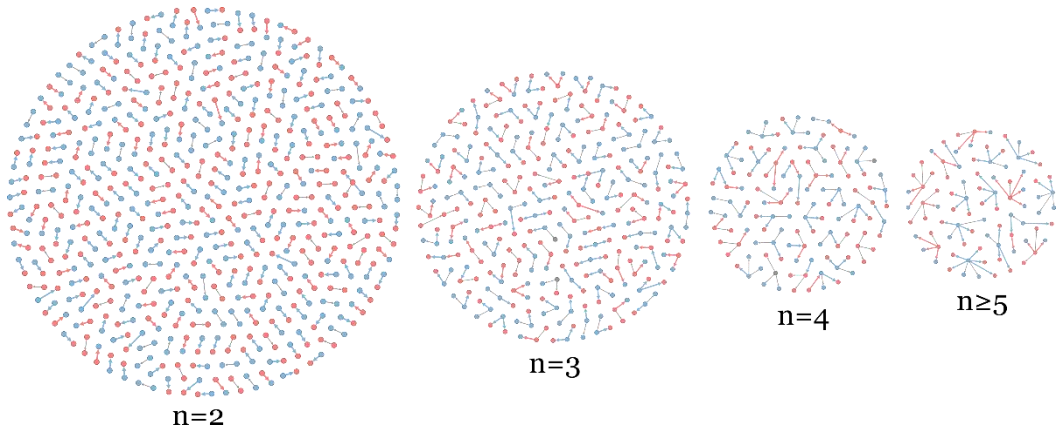

**Figure 2.** The full transmission map of considered family clusters involving 1,108 confirmed cases. The blue, red, and grey nodes represent men, women, and unknown gender cases. The blue arrows indicate transmissions with men as infectors, while the red arrows with women as infectors.

Each family cluster has only one imported case (i.e., the first-infected case). Among the 464 imported cases, 281 were men, 163 were women, and two were unknown. According to the 2020 China Statistical Yearbook, there were 5,578.34 million registered men (51.09%) and 5,340.42 registered women (48.91%) in the national resident population at the end of 2019. If gender plays no role in the transmission, the probability that a randomly sampled imported case is a man should be about 0.5109, and the number of imported male $n_m$ cases should follow a binomial distribution

$$P(n_m) = \binom{n}{n_m} q^{n_m}(1-q)^{n-n_m}, \qquad (1)$$

where $n = 464$ and $q = 0.5109$. The t-test indicates that the gender difference is significant (p<0.01), that is to say, given $n$ and $q$, the probability $P(n_m \geq 281)$ is much smaller than 0.01. Therefore, we conclude that Hypothesis 1 is verified.

**Table 1:** Statistics of the clusters with sizes 2 and 3. Blue and red nodes represent men and women, respectively.

| Transmission | Occurrence | Proportion | Transmission | Occurrence | Proportion |
|---|---|---|---|---|---|
| 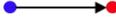 | 294 | 44.89% | 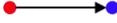 | 150 | 22.90% |
| 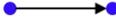 | 113 | 17.25% | 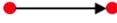 | 98 | 14.96% |
| 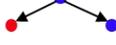 | 89 | 29.67% | 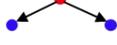 | 27 | 9.00% |
| 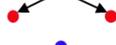 | 61 | 20.33% | 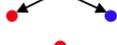 | 53 | 17.67% |
| 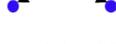 | 13 | 4.33% | 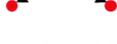 | 16 | 5.33% |
| 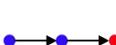 | 1 | 0.33% | 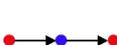 | 1 | 0.33% |
| 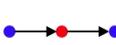 | 8 | 2.67% | 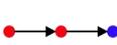 | 8 | 2.67% |
| 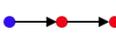 | 9 | 3.00% | 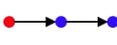 | 2 | 0.67% |
| 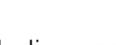 | 9 | 3.00% | 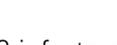 | 3 | 1.00% |

Excluding cases of unknown gender, there are 478 infectors (296 men and 182 women) and 660 infectees (266 men and 394 women). Note that a larger number of women infectees does not imply a higher probability that women get infected within families, because if the imported case in a family cluster is a man, then the proportion of women in the susceptible family members will be higher. If the null hypothesis of Hypothesis 2 holds, the probabilities of men and women being infected within the household should be the same. Accordingly, if the number of people living in a household is $S$ (only S>1 is considered) and the ratio of male infectors is $r$, then a family with male infector (at probability $r$) has in average $(qS - 1)$ susceptible men and a family with female infector (at probability $1 - r$) has in average $qS$ susceptible men, so the number of male infectees is proportional to $r(qS - 1) + (1 - r)qS$. Analogously, the number of female infectees is proportional to $r(1 - q)S + (1 - r)[(1 - q)S - 1]$. Therefore, the ratio of men to women among infectees should be

$$r_{mf} = \frac{r(qS-1)+(1-r)qS}{r(1-q)S+(1-r)[(1-q)S-1]} = \frac{qS-r}{(1-q)S-(1-r)}. \tag{2}$$

Substituting $q = 0.5109$ and $r = 296/478 = 0.6192$, we get

$$r_{mf} = \frac{0.5109S - 0.6192}{0.4891S - 0.3808}. \tag{3}$$

According to the 2020 China Statistical Yearbook, the average population per household of national residents in China is 2.92. Since the epidemic broke out during the Chinese New Year, the actual resident number of households living together should be larger than the usual value. Considering $S = 3, 4, 5$, the ratios of men to women among infectees are 0.84, 0.90, and 0.94 (the larger the $S$, the larger the ratio), respectively. All of which are significantly higher than the real situation ($r_{mf} = 0.68$). As a result, the null hypothesis is rejected, indicating that women are more likely to be infected within the household than men, namely Hypothesis 2 holds.

### 3.2 Impacts of Area and Age on Hypothesis 1

Similar to $r_{mf}$, we can use the ratio of male to female in imported cases, denoted by $r_{mf}^{imp}$, to characterize the strength of the "man as breadwinner" (MB) mode. Early studies claimed that economic development provides women with access to better-paying jobs, in particular the tertiary industry in which women enjoy comparative advantages [25,26]. We roughly divide China into three regions based on their economic developments: the east, the middle and the west. In China, the economic level in the east is higher than that in the middle, and the economic level in the middle is higher than that in the west. As shown in Fig. 3(a), $r_{mf}^{imp}$ in the three areas are all larger than 1 (suggesting the robustness of Hypothesis 1 in different areas), $r_{mf}^{imp}$ in the east is remarkably larger than $r_{mf}^{imp}$ in the middle, and $r_{mf}^{imp}$ in the middle is remarkably larger than $r_{mf}^{imp}$ in the west. That is to say, the MB mode seems to be stronger in a more economically developed area, indicating that the development of economics in current China may not reduce gender difference in career, which is consistent with the observations reported in [27].

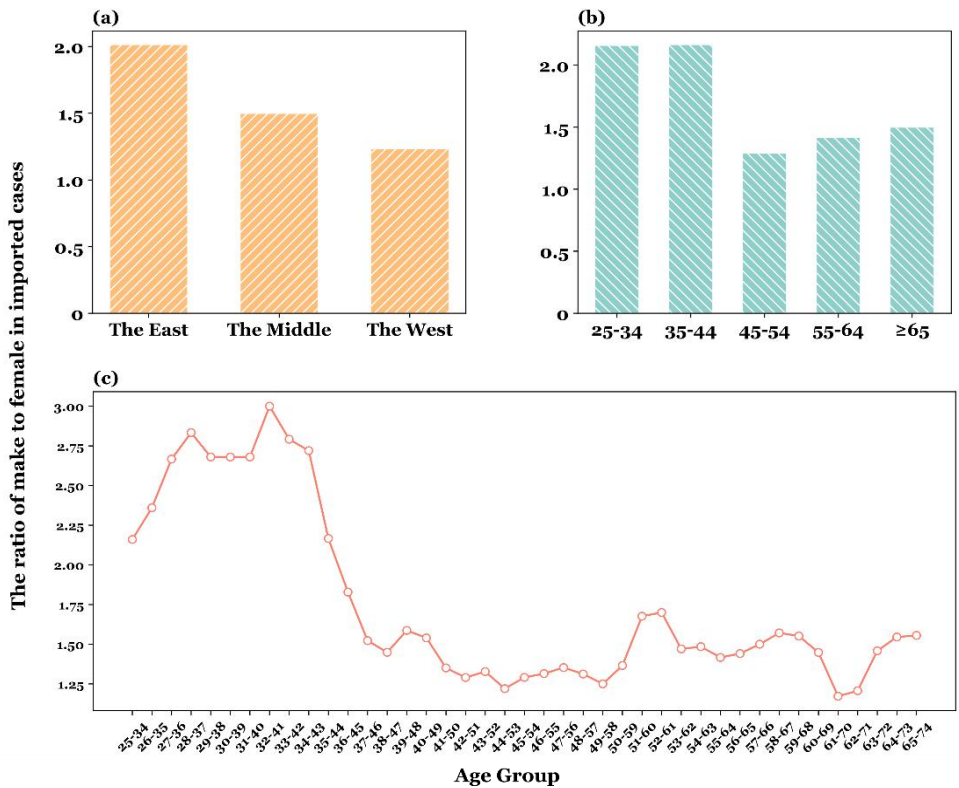

**Figure 3.** The ratios of imported male cases to imported female cases in subgroups: (a) grouped by three geographical areas in mainland China, (b) grouped by non-overlapped age bins, (c) grouped by overlapped age bins.

We next divide imported cases with ages ≥25 into 5 groups: 25-34, 35-44, 45-54, 55-64, ≥65. As shown in Fig. 3(b), $r_{mf}^{imp}$ is larger than 1 for every group, and $r_{mf}^{imp}$ for younger group is larger than that for elder group. Figure 3(c) further shows the $r_{mf}^{imp}$-age correlation, where imported cases with ages≥25 are divided into overlapped groups each covering 10 years. Analogous to the non-overlapped case, $r_{mf}^{imp}$ is negatively correlated to the age and the younger generation has remarkably larger $r_{mf}^{imp}$. Such negative correlation may simultaneously caused by multiple reasons, such as the observed gradually expanding gender gap in China's workplace [28,29] and the motherhood penalty suffered by young women [30,31]

**3.3 Impacts of Family Role on Hypothesis 2**

Aiming at analyzing the robustness of Hypothesis 2, we stratify the transmissions by gender and age groups. Among the 662 family-clustered transmissions, 31 involved individuals of unknown gender or age. Thus we eliminate these 31 transmissions and create a bipartite network based on the remaining 631 transmissions. We simply divide the confirmed cases into three generations according to their ages: the first generation (M1 and F1, 0-24), the second generation (M2 and F2, 25-49), and the third generation (M3 and F3, ≥50). As shown in Fig. 4(a), a weighted bipartite network can be created. The nodes on the left represent the infectors, and the nodes on the right represent the infectees. A few individuals are both infectors and infectees. Edges represent transmissions. For example, the edge "M3 → F3" represents transmissions from men over 50 years old to women over 50 years old.

We established a null model that assumes the gender plays no role and accounts for the generation structure of families. This can be treated as a variant of the null hypothesis of Hypothesis 2, which still rejects the gender role but assumes that the observed gender difference in infectees may be resulted from the different transimission patterns between different generation pairs. The null model follows the three rules below. (1) The distribution of infectees in the three generations only depends on the infector's generation, not gender. This distribution is statistically consistent with Fig. 4(b). Among the 631 family-clustered transmissions, 239 occurred between the same generation, 348 occurred between neighboring generations, and 44 crossed two generations. (2) An infector only infects the opposite gender in the same generation, as the transmissions between the same gender in the same generation have only 13 occurrences, accounting for merely 2.06% of all transmissions, which can be ignored. (3) If a transmission happens between different generations, the gender of the infectee is randomly chosen.

Figure 4(c) shows the result of the null model. The difference between the number of male infectees and the number of female infectees in the third generation is 75 (173-98) in the real data but only 21 (146-125) in the null model. Analogously, the difference in the second generation is 66 in the real data but only 34 in the null model. Consequently, even if we consider the generation structure, the null hypothesis of Hypothesis 2 is rejected according to the significant differences between Figs. 4(a) and Fig. 4(c).

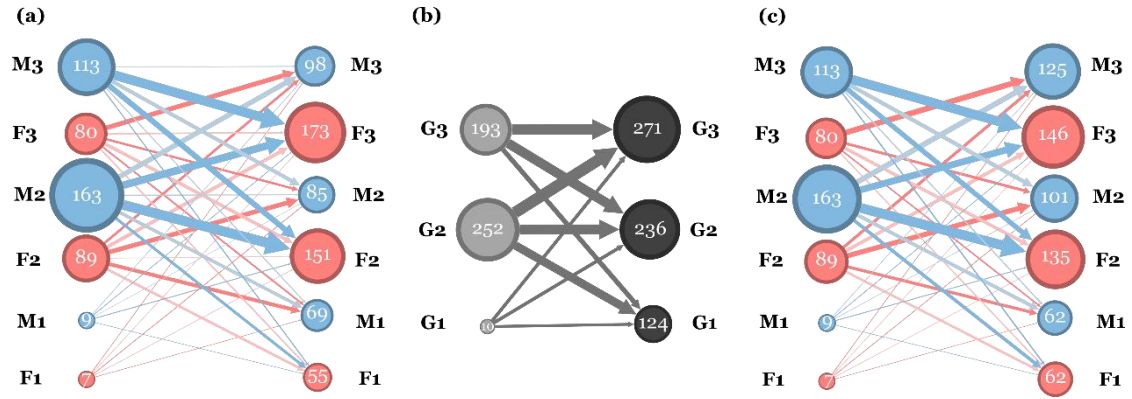

**Figure 4.** Family-clustered transmissions grouped by (a) age and gender, (b) age only, and (c) age and gender in the null model. Blue and red nodes represent men and women; a node's size is proportional to the number of cases. In plot (a) and plot (c), M1 – M3 and F1 – F3 are the first to third generations of men and women, respectively. Edges with darker colors represent transmissions between different genders, while those with lighter colors are between the same gender. In plot (b), G1, G2, and G3 represent the first, second, and third generations.

We further analyze the impacts of specific family relationships. According to the social norm in China, we consider four types of family relationships, namely husband-wife (HW), parent-child (PC), spouse-parent-child (SPC), and grandparent-grandchild (GG) relationships. Figure 5 shows the transmission networks with edges representing different family relationships. Next, we distribute transmissions resulted from the null model (i.e., the model that generates Fig. 4(c)) to different family relationships, assuming that a couple has the same probability of staying with each's parents. For example, to calculate the number of the mother → son transmissions, we only need to consider F3 → M2 (the corresponding edge weight in the null model is 22.6) and F2 → M1 (the corresponding edge weight in the null model is 15.7). As F3 → M2 also contains mother-in-law → son-in-law transmissions, the weight should be divided by 2. Therefore, the approximate number of mother → son transmissions is 22.6/2+15.7=27.0. The relative difference between actual data and the null model is defined as

$$r_d = \frac{n_r - n_0}{n_r + n_0}, \qquad (4)$$

where $n_r$ is the actual number of cases and $n_0$ is the corresponding number in the null model. When $r_d > 0$, the actual number is larger than the predicted number by the null model that accounts for the family relationships, while when $r_d < 0$, the actual data is smaller.

Table 2 presents the number of transmissions and relative differences for all considered family relationships. The six relationships ranked by $r_d$ in descending order and with $n_r > 10$ are son → mother, mother → son, son → father, mother-in-law → daughter-in-law, daughter → mother, and mother → daughter. This result demonstrates again that women are more likely to be infected than men within the household.

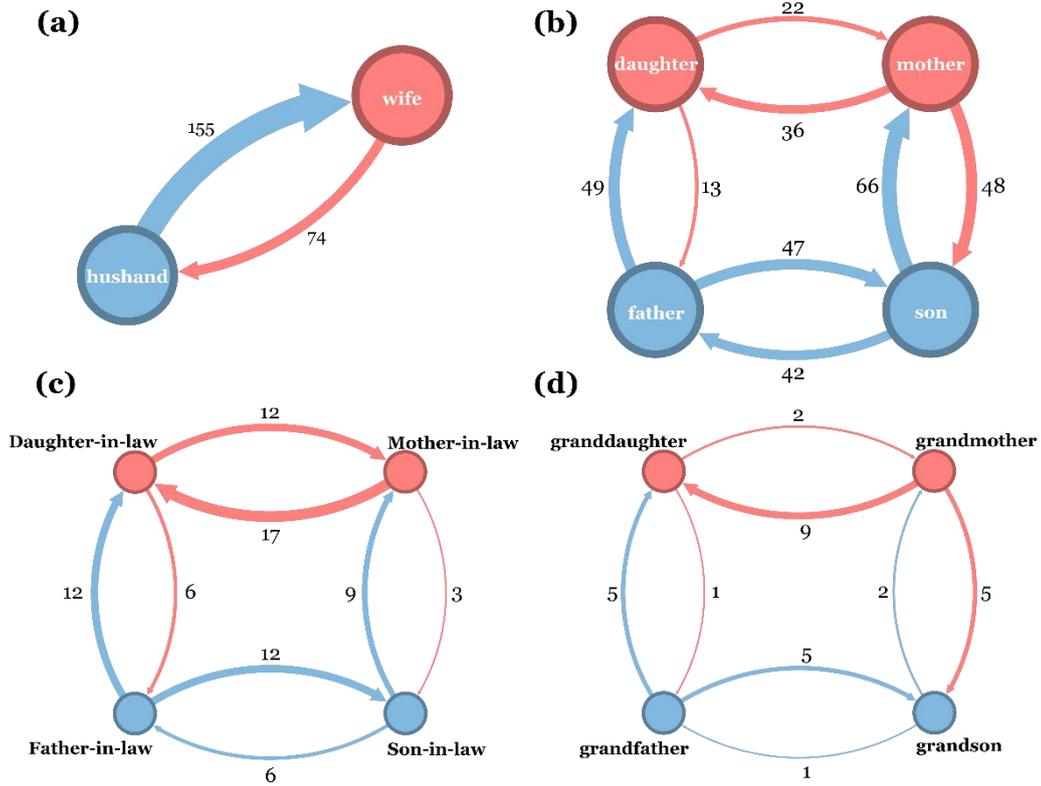

**Figure 5.** Transmissions among different family relationships: (a) husband-wife (HW) relationship, (b) parent-child (PC) relationship, (c) spouse-parent-child (SPC) relationship, (d) grandparent-grandchild (GG) relationship. A directed edge from A to B indicates A infects B, with the associated number (edge weight, proportional to the edge thickness) being the number of corresponding transmissions.

**Table 2**: The number of transmissions in the data ($n_r$) and the null model that accounts for family relationshuo ($n_0$) for different family relationships, as well as the corresponding relative differences $r_d$.

| Generations | Relationships | $n_r$ | $n_0$ | $r_d$ |
|---|---|---|---|---|
| G3 → G3 and G2 → G2 | husband → wife | 155 | 145.88 | 0.0303 |
| | wife → husband | 74 | 91.12 | -0.1037 |
| G3 → G2 and G2 → G1 | father → son | 47 | 44.74 | 0.0246 |
| | father → daughter | 49 | 44.74 | 0.0454 |
| | mother → son | 48 | 27.01 | 0.2798 |
| | mother → daughter | 36 | 27.01 | 0.1427 |
| G2 → G3 and G1 → G2 | son → father | 42 | 25.57 | 0.2432 |
| | son → mother | 66 | 25.57 | 0.4415 |
| | daughter → father | 13 | 14.68 | -0.0607 |
|  | daughter → mother | 22 | 14.68 | 0.1996 |
| G3 → G2 | father-in-law → son in law | 12 | 15.95 | -0.1413 |
|  | mother-in-law → son in law | 3 | 11.30 | -0.5804 |
|  | father-in-law → daughter in law | 12 | 15.95 | -0.1413 |
|  | mother-in-law → daughter in law | 17 | 11.30 | 0.2014 |
| G2 → G3 | son in law → father-in-law | 6 | 22.48 | -0.5787 |
|  | son in law → mother-in-law | 9 | 22.48 | -0.4282 |
|  | daughter in law → father-in-law | 6 | 12.27 | -0.3432 |
|  | daughter in law → mother-in-law | 12 | 12.27 | -0.0111 |
| G3 → G1 | grandfather → grandson | 5 | 9.66 | -0.1554 |
|  | grandfather → granddaughter | 5 | 9.66 | -0.3179 |
|  | grandmother → grandson | 5 | 6.84 | -0.3179 |
|  | grandmother → granddaughter | 9 | 6.84 | 0.1364 |
| G1 → G3 | grandson → grandfather | 1 | 3.09 | -0.5110 |
|  | grandson → grandmother | 2 | 3.09 | -0.2142 |
|  | granddaughter → grandfather | 1 | 2.41 | -0.4135 |
|  | granddaughter → grandmother | 2 | 2.41 | -0.0930 |

## Discussion

By analyzing the spreading pattern in family clusters of the COVID-19 epidemic, this paper reveals the gender inequality in care work from a novel perspective and provides novel supportive evidence for the existence of the "men as breadwinner and women as homemaker" (MBWH) mode in China. Further analyses show that: (1) the MB mode is robust when considering area-specific or age-specific subgroups, (2) the MB mode is stronger for better economically developed area and younger generation, and (3) the WH mode is robust when family roles are taken into consideration. This paper applies null models to validate the statistical significance. The usage of null models becomes increasingly popular in natural science, as null models can be flexibly designed with target variables being well controlled in highly complicated situations [32,33]. In comparison, this tool is rarely used in social science, and we hope social studies with complicated statistics can draw on the experience of the usage of null modes in this paper.

It is widely acknowledged that unequal care work is a critical factor that leads to gender inequality [34]. Having the traditional culture that admires the MBWH mode, the housework gap between men and women is significant in China. According to the *National Time*

*Utilization Survey* released by the National Bureau of Statistics of China in 2018, Chinese married men spend an average of 81 minutes per day on housework (including taking care of children and elderly relatives), while married women spend as much as 225 minutes. Recent analysis shows that this difference may play a more important role than the difference in human capital (mainly resulting from gender inequality in education) in forming the gender pay gap [27].

According to COVID-19 pathological studies, there is no significant physiological difference between men and women in susceptibility [35]. However, as shown in this paper, the spreading pattern is affected by the gender division of labor. The MBWH mode makes men more likely to become imported cases and women more likely to be infected within the household. In particular, contacts between family members are close and frequent and thus difficult to be controlled through social distancing policies. For example, during the outbreak period, contacts that happened outside decreased sharply after the lockdown, and most contacts (about 78%) occurred between family members [36]. In family-clustered cases, transmissions among different gender groups happen with different probabilities. This reminds us to adopt gender-differentiated epidemic prevention measures. For example, in the case of MBWH mode unchanged, women need to be prioritized to be trained to effectively take care of confirmed or suspected cases and avoid infection. This is not only for the COVID-19 epidemic, but also for other infectious diseases. The perspective and method of gender-differentiated analyses in this paper can also be used and extended to analyze human groups of different ages, races, and socioeconomic status, aiming at designing human-oriented epidemic prevention policies.

An obvious limitation in this study is the small size and insufficient dimensions of the data. Firstly, because of the small number of family-clustered cases, we cannot provide solid statistics about the relationships between the strength of the MB mode and the geography or economics, or reveal clear picture about different transmission patterns through different family relationships. Secondly, because of the shortage of detailed information of the social and economics states of those involved families, we cannot build up trustable causal relationships among potentially related factors.

## ACKNOWLEDGMENTS

This work was partially supported by the National Natural Science Foundation of China (Grant Nos. 11975071 and 62173065), the Science Strength Promotion Programmer of the University of Electronic Science and Technology of China under Grant No. Y03111023901014006, and the Fundamental Research Funds for the Central Universities of China under Grant No. ZYGX2016J196.

## AUTHOR CONTRIBUTIONS

J.L., X.F.L., X.X., and T.Z. conceived and designed the project, X.F.L. and X.X. collected the data, J.L. performed the research, J.L., X.F.L., X.X., and T.Z. analyzed the data, J.L. and T.Z. wrote the manuscript, X.F.L. and X.X. edited the manuscript.

# DECLARATION OF INTERESTS

The authors declare no competing financial interests.